# A physical model predicting instability of rock slopes with locked segments along a potential slip surface

Chen Hongran[a, b, c], Qin Siqing[a, b, c*], Xue Lei[a, b, c], Yang Baicun[a, b, c], Zhang Ke[a, b, c]

a Key Laboratory of Shale Gas and Geoengineering, Institute of Geology and Geophysics, Chinese Academy of Sciences, Beijing 100029, China

b Institutions of Earth Science, Chinese Academy of Sciences, Beijing 100029, China

c University of Chinese Academy of Sciences, Beijing 100049, China

*Corresponding author at: Key Laboratory of Shale Gas and Geoengineering, Institute of Geology and Geophysics, Chinese Academy of Sciences, Beijing 100029, China. Tel. /fax: +86 010 82998616.

E-mail address: qsqhope@mail.iggcas.ac.cn (Qin. S. Q).

**Abstract**

Predicting the occurrence of landslides is important to prevent or reduce loss of lives and property. The stability of rock slopes is often dominated by one or more locked segments along a potential slip surface; these segments have relatively high strength and accumulate strain energy. Locked segments can be preliminarily classified into three categories: "rock bridge", "retaining wall" and "sustaining arch." Coupling a one-dimensional renormalization group model with the strain-softening constitutive relation of geo-materials considering the Weibull's distribution, a physical model for predicting the instability of slopes with locked segments is established. It is found that the ratio of the strain or displacement at the peak strength point to that at the volume dilation point for a locked segment is exclusively dependent on the Weibull's shape parameter $m$, and is approximately constant at 1.48. A corresponding accelerating displacement increment (tertiary creep) of the slope can be observed from the onset of the volume dilation of the locked segment due to its unsteady cracking. For a slope with multiple locked segments, one can predict its critical instability displacement value according to the accelerating displacement onset corresponding to the volume dilation point of the first locked segment and the number of locked segments. The back-analysis of two typical cases, the Yanchihe rockslide in China and the wedge rockslide, Libby Dam, USA, shows that their evolutionary processes, dominated respectively by one and two locked segments, follow this model, confirming the reliability of the proposed model.

## 1. Introduction

A landslide is the movement of a mass of rock, debris, or soil down a slope under the influence of gravity (Cruden, 1991). Landslides cause significant damage and casualties every year. Brabb (1993) claimed that at least 90% of landslide damage can be avoided if the risk is recognized before the landslide event. Hence, in an effort to develop a reliable method for landslide prediction researchers have used a variety of approaches, which primarily fall into two categories: phenomenological and physical approaches.

Phenomenological approaches include empirical and regression-only methods. Empirical methods often derive from the pre-failure accelerating phase of the strain–time (or displacement–time) creep curve. Saito (1965) performed the first successful prediction using a model where the time to failure in the tertiary creep phase was inversely proportional to the existing strain rate. On the basis of Saito's model, Fukuzono (1985) introduced an extended model which is expressed as the inverse-velocity of the displacement. The inverse-velocity model has been improved and extensively utilized (Voight, 1988; Crosta and Agliardi, 2003; Sornette et al., 2004; Rose and Hungr, 2007; Mufundirwa et al., 2010) because of its simplicity. Compared to the empirical methods mentioned above, regression-only methods rely on regression functions, such as fuzzy logic (Champatiray et al., 2006) and artificial neural networks (Mayoraz and Vulliet, 2002; Du et al., 2013), to represent complex correlations among several landslide-triggering factors. Generally speaking, phenomenological approaches do not take into account the mechanical behavior of the geological body or its boundary conditions. The lack of any specific relation with the physics of the phenomenon would make the predicting results mainly of academic interest, and the accuracy of the predicted time of the event seems to be merely a matter of coincidence (Federico et al., 2012). In contrast, the physical approach based on the landslide mechanism promises more reliable landslide prediction results.

In many reported physical models based on limited equilibrium analysis, the sliding body is simplified as an infinite block on a continuous inclined plane. The movement of the slope can be determined with the kinematic (Herrera et al., 2009), dynamic (Yalçınkaya and Bayrak, 2003) or momentum equation (Corominas et al., 2005) where the constitutive relations of the geo-materials, rainfall and hydraulic models are taken into account; such models include the viscoelastic model (He et al., 2005), rainfall spatial distribution (Chen and Zhang, 2014) and subsurface flow model (An et al., 2016). When a certain criterion is reached, the block will slide along the plane, representing the slope instability. Physical models consider geological factors when estimating slope stability, but the hypothesis of a continuous slip surface often contributes to inaccurate representation of the mechanism of some landslides.

Lajtai (1969) pointed out that potential slip surfaces are usually discontinuous. Intact "rock bridges" are commonly found between joints that constitute the slip surface; slope failure occurs when the rock bridges reach their shear strength (Eberhardt et al., 2004). The failure probability of an unstable rock mass depends mainly on the proportion of rock bridges (Frayssines and Hantz, 2006) along the potential slip surface. Any part of a slope that governs the stability of the slope can be defined as a "locked segment" (Qin et al., 2010a; Huang, 2015). Thus, the locked segments are the key to the analysis of progressive failures in many slopes. For example, in the Jiweishan rockslide in Chongqing, Southwestern China, the movement of the upper rock blocks was restricted by a lower rock block (Tang et al., 2015), i.e., a locked segment. When the locked segment failed, massive blocks of rock slid subsequently. Unfortunately, this mechanism was not clearly recognized before the slope failure, and the volume (~5 million m$^3$) and travel distance (~2.2 km) of the sliding masses were underestimated. In less than 1 min, 74 people were killed in the rockslide (Xu et al., 2010). This example shows that complete failure of the locked segment usually produces large-scale and high-speed landslides, leading to huge losses in lives and property, and emphasizes the significance of research on the damage mechanism of locked segments.

Hence, a growing number of studies have focused on the locked segment. Pan et al. (2014)

analyzed the formation mechanism of locked segments observed in a number of landslide cases. Laboratory tests of physical models (Huang et al., 2016) demonstrated that macroscopic failure of locked segments resulted in high-speed rock slides and that the failure process was influenced by the location (Huang et al., 2015), number and length of the locked segments and the distance between them (Pan et al., 2017).

However, our understanding of both the geological characteristics and mechanical behavior of locked segments is still lacking, the geological characterization of the locked segments is not systematic, and predicting the instability of slopes with locked segments is still difficult. The failure of a locked segment is accompanied by abrupt and intensive release of energy, which is occasionally characterized by nearly instantaneous displacements in short time intervals. The inverse-velocity method cannot be used to model such a process (Rose and Hungr, 2007; Federico et al., 2012). A locked segment may not fail after heavy rainfall; therefore, false alarms may be issued based on rainfall threshold estimates. Some physical models based on continuum mechanics (Lajtai, 1969; Einstein et al., 1983), fracture mechanics (Kemeny, 2005) and safety factors (Huang et al., 2015) were established to describe the rupture of rock bridges, but their complicated expressions contain variables, such as the friction coefficient and critical strength of the rock, which are difficult to measure. Currently, no specific approaches have been universally accepted to predict landslides with locked segments.

In recent years, a physical prediction model (Qin et al., 2010a; Qin et al., 2010b; Xue et al., 2014a; Xue et al., 2017), which couples a renormalization group model with the constitutive relation based on Weibull's distribution, was established. Back analysis showed that this model is promising for predicting the instability of slopes with locked segments. In this paper, based on analysis of reported landslide cases as well as large-scale slopes that had been investigated in detail, a systematic classification of locked segments is conducted, and then a model for rock slopes with one or multiple locked segments is introduced with a specific limit placed on Weibull's shape parameter. This work will help to better understand the failure mechanism of slopes with locked segments, and may provide guidelines for disaster mitigation and prevention.

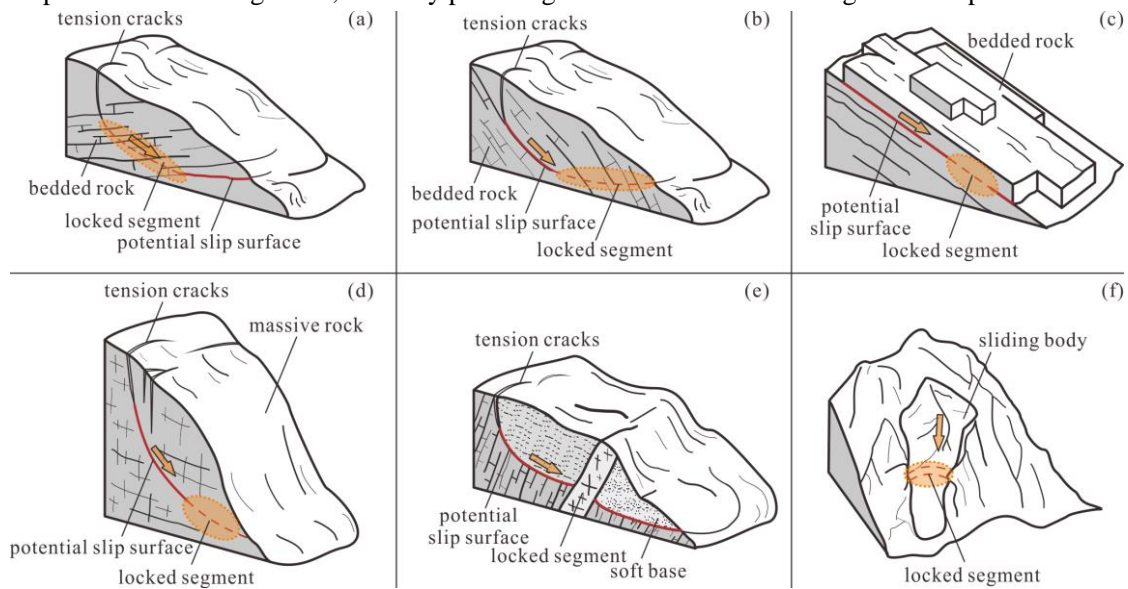

Fig. 1 Primary categories of locked segment types. (a)–(d) Various "rock bridges", (e) "retaining wall" and (f) "sustaining arch."

## 2. The geological categorization of locked segments

Locked segments may be categorized into various types depending on the geological conditions of the slopes, such as the geomorphology, structure and lithology. Without clarifying these complex characteristics and categorizing the locked segments, it is difficult to correctly identify slopes with locked segments and understand their mechanisms of instability; thus, the prediction of landslides in slopes with locked segments is currently unreliable.

Here we summarize the types of locked segments for many typical landslides and slopes (Wang et al., 1988; Cheng et al., 2004; Pan et al., 2014; Huang, 2015; Pan et al., 2017; Xue et al., 2017) in terms of engineering geology, and classify them into three categories (Figure 1): "rock bridge," "retaining wall" and "sustaining arch."

Generally speaking, the lithology of a rock bridge is identical with the surrounding lithology. Rock bridges can be subdivided into the following three types according to their geological characteristics: (i) in a stratified slope, a potential slip surface usually intersects with the layered strata which play the role of the locked segment, such as an anti-dip stratified slope (Figure 1a) and a dip stratified slope whose dip angle is larger than the slope angle (Figure 1b); (ii) in a stratified slope which tends to move along a bedding plane, the sliding is usually controlled by an intact rock bridge, i.e., a locked segment on the bedding plane (Figure 1c); (iii) when a slope consists of massive rock without distinct bedding planes, as illustrated in Figure 1d, the locked segment is similar to a relatively homogeneous rock bridge.

When a hard stratum occurs in the middle of a slope, it acts as a locked segment, preventing the upper part of the slope mass from moving and acting like a "retaining wall" (Figure 1e). Once the locked segment fails, the soft lower part is unable to resist the movement of the upper block, leading to slope instability.

The "sustaining arch" mechanism (Figure 1f) was first presented in the study of the Xintan landslide (Wang et al., 1988). The relatively narrow geomorphology in the middle of the Xintan slope led to a local zone of stress enhancement where a sustaining arch structure was formed. The landslide occurred when the arch structure failed. Any part of the slope that produces this effect, such as the huge boulders accumulating in the Houziling slope body (Cheng et al., 2004), can be viewed as the locked segment.

The classification of locked segments provided clear geological models for our research on the physical model for prediction of landslides with locked segments.

## 3. Relationship between a typical creep curve and the deformation and failure process of the locked segment

A typical creep curve of a rock specimen usually includes the following stages (Figure 2a): (1) primary creep (MR), (2) secondary creep (RQ) and (3) tertiary (accelerating) creep (QS). Creep acceleration can be observed from the onset of the tertiary stage (point Q), leading to macro fracture of the specimen (point S). Since a locked segment affected by various environmental factors undergoes a deformation and failure process similar to that of the rock specimen, the displacement–time creep curve of a rockslide controlled by a single locked segment should have a similar shape to that of a typical creep curve.

Experimental investigations (Bieniawski, 1967; Martin and Chandler, 1994; Xue et al., 2014b; Xue et al., 2015a) have shown that the ideal deformation and failure process of rock samples subjected to compressive or shear loading can be divided into five stages (Figure 2b): (I) crack closure (OA), (II) elastic deformation (AB), (III) stable crack growth (BC), (IV) unstable crack growth (CD), and (V) post-peak failure (DE). When the loading exceeds the damage stress threshold at point C, the cracks propagate in an unstable manner, resulting in volume dilation (Brace et al., 1966; Martin and Chandler, 1994), which cannot be restrained even when the load is held constant. For a rock sample under creep loading, the damage during the tertiary creep stage depends on the dilatant volume formed by unstable micro cracking (Kranz and Scholz, 1977; Reches and Lockner, 1994). From the tertiary creep onset, the sum of acoustic emission (AE) representing cracking increases rapidly, and macro cracks that penetrate the sample begin to form (Figure 3) (Sun, 1999). Xue et al. (2014a) pointed out that point C on the stress–strain curve corresponds to point Q on the creep curve. Likewise, as illustrated in Figure 2, the time-dependent strain (displacement) curve of a locked segment can be associated with the stress–strain curve of strain-softening geo-material.

Usually, the intensity of the brittle failure of a locked segment such as a rock bridge is so strong that the stress drop during the post-peak stage is quiet large in a short period, implying that for a rock slope with a locked segment the anti-sliding force may be far less than the driving force after the locked segment fails. Especially for slopes whose potential slip surface is steep, the instability of the slope is likely to take place soon after the formation of the transfixion surface. Hence, for these slopes point S, i.e., the peak strength point D will approximately represent the ultimate landsliding point.

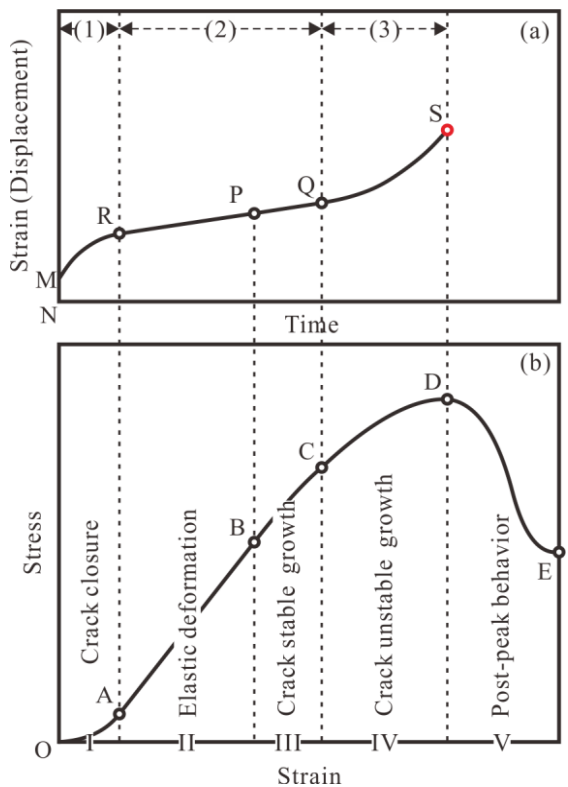

Fig. 2 (a) Typical creep curve and (b) the deformation and failure process of a rock specimen or locked segment subjected to compressive or shear loading. The symbols (1), (2) and (3) in (a) indicate the primary, secondary and tertiary creep stages, respectively.

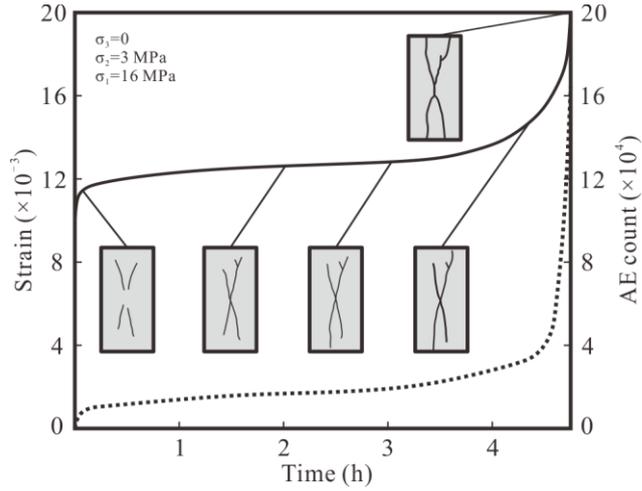

Fig. 3 Evolution of strain (solid curve) and AE counts (dotted curve) with time for a jointed rock sample under biaxial compression. The five sketches of the fracture patterns show the observed progressive failures of the sample (modified after Sun (1999)).

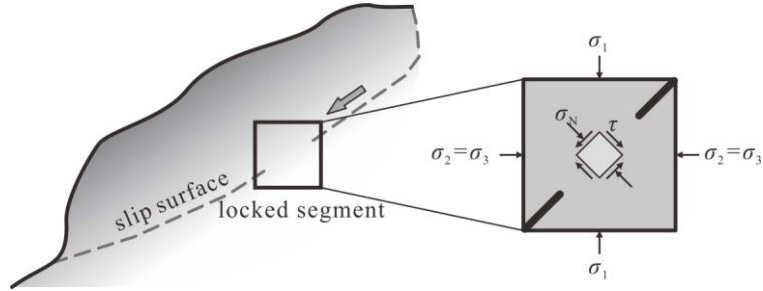

Fig. 4 Schematic diagram of stress state in a locked segment subjected to compressive and shear loading along a potential slip surface. In the right zoomed-in view of the locked segment, $\sigma_1$, $\sigma_2$ and $\sigma_3$ are the maximum, medium and minimum principal stresses; $\sigma_N$ is the normal stress and $\tau$ is the shear stress.

## 4. Prediction model

The locked segment can be assumed to be composed of numerous cells, the smallest indivisible units, whose strains $\varepsilon$ follow Weibull's distribution (Weibull, 1951). Thus, the failure probability of the cells ($p_0$) proposed by Qin et al. (2006) is

$$p_0 = 1 - \exp\left(-\frac{\varepsilon}{\varepsilon_0}\right)^m \tag{1}$$

where $\varepsilon_0$ is the measured mean strain and $m$ is the Weibull's module (shape parameter).

The general compressive applied state of stress for a locked segment (Figure 4), such as a rock bridge, can be reduced to a state of direct shear with a nearly fixed normal compressive stress (Lajtai, 1969). In this state the constitutive equation between the shear stress $\tau$ and strain $\varepsilon$ of the locked segment along an unconnected slip surface based on Eq. (1) (Qin et al., 2010a; Qin et al., 2010b; Xue et al., 2014a) can be determined by

$$\tau = G \cdot \varepsilon \cdot \exp\left[-\left(\frac{\varepsilon}{\varepsilon_0}\right)^m\right] \tag{2}$$

where $G$ is the shear module of the locked segment. The shape of the function curve is controlled by the module $m$: the higher the $m$-value, the steeper the curve (Figure 5).

To attain the expression of the stress peak at point D (Figure 2b), the first-order derivative of the constitutive equation (1) can be set equal to zero:

$$\frac{\mathrm{d}\tau}{\mathrm{d}\varepsilon} = G \cdot \exp\left[-\left(\frac{\varepsilon}{\varepsilon_0}\right)^m\right] \cdot \left[1 - m \cdot \left(\frac{\varepsilon}{\varepsilon_0}\right)^m\right] = 0 \qquad (3)$$

substituting $\varepsilon_D$, the strain at point D, into Eq. (3) leads to

$$\frac{\varepsilon_D}{\varepsilon_0} = \left(\frac{1}{m}\right)^{\frac{1}{m}} \qquad (4)$$

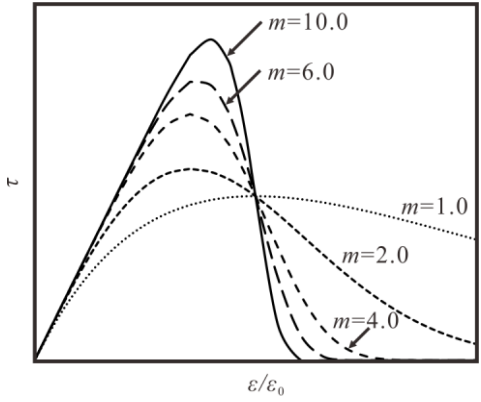

Fig. 5 Curves of the constitutive relation for different *m*-values.

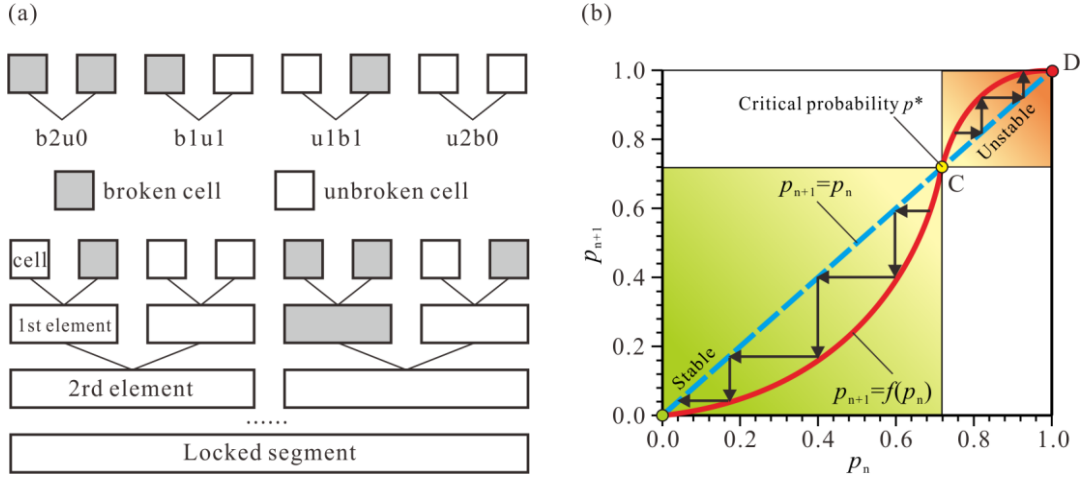

Fig. 6 (a) Illustration of the one-dimensional renormalization group (RG) model. (b) The dependence of the failure probability for the (*n*+1)th order element on the failure probability of the *n*th order element. The critical failure probability (*p*\*) gives the bifurcation point C for catastrophic failure of the locked segment at point D (modified after Xue et al. (2014a)).

Point C is a characteristic point corresponding to the volume dilation onset, indicating that the locked segment is in the critical state where cracks tend to grow unsteadily. This is equivalent to the concept of a critical point (Xue et al., 2015b) in Renormalization Group (RG) theory (Wilson, 1971) which has been applied to analysis of the critical state of the rock fracture process (Smalley et al., 1985; Qin et al., 2010a; Qin et al., 2010b).

The state of each cell that constitutes the locked segment is either broken (grey) or unbroken (white) (Figure 6a). Provided that two cells constitute a first-order element, the element state (broken or not) is dependent on the cell combination. There are four combinations of two cells in the one-dimensional RG model: b2u0, b1u1, u1b1, and b0u2. A first-order element is defined as broken when both its sub-cells are broken; therefore, the failure probability of the first-order element $p_1$ (Qin et al., 2010a; Qin et al., 2010b; Xue et al., 2014a) is

$$p_1 = 2p_0 - p_0{}^2 - 2p_0(1 - p_0)^{2^m} \qquad (5)$$

For an element of the ($n$+1)th order, the failure probability follows the same rule (Figure 6a); thus, Eq. (5) is written in a general form as

$$p_{n+1} = 2p_n - p_n^2 - 2p_n(1-p_n)^{2^m} \qquad (6)$$

When the failure probability exceeds $p^*$, corresponding to the critical point on the S-shaped curve (Figure 6b) of Eq. (6), the failure probability of the locked segment will spontaneously approach 1.0, indicating that the state of cracking within the locked segment changes from stable to unstable (Qin et al., 2010a; Qin et al., 2010b; Xue et al., 2014a). The solution of $p^*$ can be obtained by letting $p_{n+1}=p_n$ in Eq. (6):

$$p^* = 1 - 2^{\frac{1}{1-2^m}} \qquad (7)$$

The failure probability of the cells at the volume dilation point C from Eq. (1) is

$$p^* = 1 - \exp\left(-\frac{\varepsilon_C}{\varepsilon_0}\right)^m \qquad (8)$$

Combining Eqs. (7) and (8) leads to the mathematical description of point C:

$$\frac{\varepsilon_C}{\varepsilon_0} = \left(\frac{\ln 2}{2^m-1}\right)^{\frac{1}{m}} \qquad (9)$$

Because $\varepsilon_0$ is inconvenient to measure, we cannot obtain the actual strain at point C and D in advance. Fortunately, $\varepsilon_0$ can be eliminated in the expression of the ratio of Eq. (4) to (9). Since the value of $\varepsilon_C$ can be determined from the accelerating strain onset, the critical instability strain $\varepsilon_D$ according to the ratio can be predicted by

$$\frac{\varepsilon_D}{\varepsilon_C} = \left(\frac{2^m-1}{m\ln 2}\right)^{\frac{1}{m}} \qquad (10)$$

This is the quantitative strain relationship between the volume dilation point and peak strength point. The ratio $\varepsilon_D/\varepsilon_C$ is exclusively dependent on the $m$-value. The parameter $m$ represents both the heterogeneity of the rock and its mechanical response to the loading conditions, and is an index generally describing the intensity of brittle failures (Chen et al., 2017). For strain-softening geo-material, the $m$-value is usually not less than 1.0. The theoretical relation between the $m$-value and the fractal dimension ($D_f$) of the crack distribution (Yang et al., 2017) is expressed as

$$m = 2D_f \qquad (11)$$

Because the range of $D_f$ is (0, 3.0), the corresponding $m$-value is limited to within [1.0, 6.0]. A recent study (Chen et al., 2017) shows that for a large rock specimen (locked segment) usually with a small height-to-diameter ratio and subjected to shear loading at an extremely low strain rate, the reasonable range limit of the $m$-value is [1.0, 4.0]. Within this range, the ratio of $\varepsilon_D/\varepsilon_C$ is insensitive to variations in the $m$-value (Figure 7) and hence can be approximately expressed by the constant 1.48, the average value of $\varepsilon_D/\varepsilon_C$ within [1.0, 4.0]. Thus, for a slope with a single locked segment, its critical strain at instability is

$$\varepsilon_D = 1.48\varepsilon_C \qquad (12)$$

Equation (12) can also be written as the sliding displacement along a slip surface:

$$u_D = 1.48u_C \qquad (13)$$

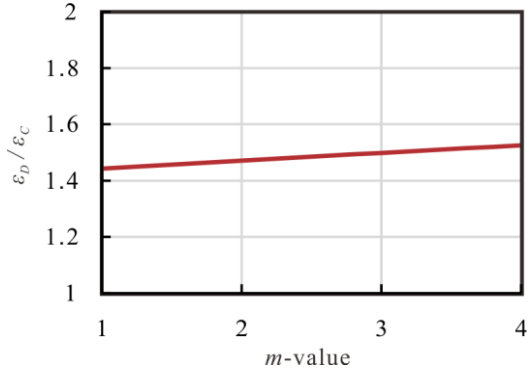

Fig. 7 Relation between $\varepsilon_D/\varepsilon_C$ and $m$.

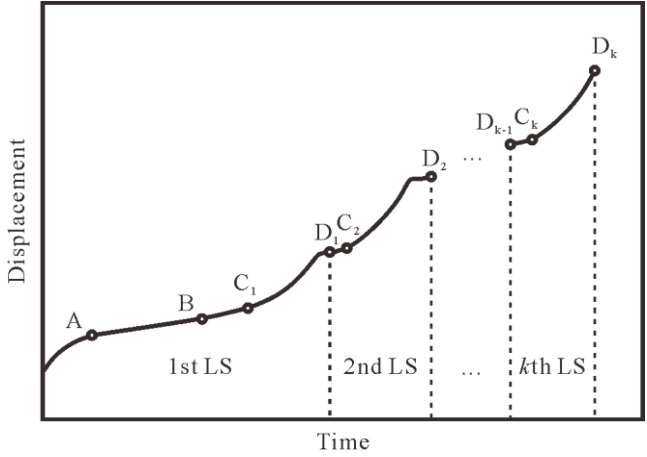

Fig. 8 Typical displacement–time curve of a slope with $k$ locked segments (LS) along a potential slip surface.

This is the prediction model for a slope with only one locked segment. However, there may be more than one locked segment along the potential slip surface of a natural slope. The evolution of a slope with $k$ locked segments involves the sequential rupture of locked segments, from the weakest to the strongest one. When the first locked segment is damaged to reach its volume dilation point, the corresponding first-time accelerating creep of the slope (point $C_1$ in Figure 8) begins and the locked segment quickly loses the capacity to resist the slide of the slope because of its strength degradation. Accordingly, the second locked segment will bear a larger driving force due to the added stress transferred to it. When the driving force and anti-sliding force of the slope come to a new equilibrium state, the movement of the slope is restricted again. However, the movement resumes under the influence of external environmental factors such as rainfall; eventually, the first locked segment will fail at point $D_1$. Subsequently, when the second segment is damaged to reach its volume dilation point $C_2$, the corresponding second stage of accelerating creep of the slope will start. As the locked segments fail one by one, the final instability of the slope occurs once the last locked segment fails. Hence, the displacement–time curve of the slope is characterized by a step-shaped profile (Figure 8), with the number of the steps being dependent on the number of locked segments.

To introduce the model with multiple locked segments we take a slope with two locked segments as an example. Because the first locked segment has lower strength than the second one, it will fail first. The relationship between the displacement at the peak strength point ($u_{D1}$) and that at the volume dilation point ($u_{C1}$) of the first locked segment is expressed in a similar form to Eq. (13):

$$u_{D1} = 1.48u_{C1} \qquad (14)$$

Once it is broken, the shear stress will be totally transferred to the second locked segment. After an increment of displacement, $\Delta u$, the volume dilation point of the second locked segment is reached. Similarly, the ratio of the second locked segment, $u_{D2}/u_{C2}$, can be expressed as

$$u_{D2} = 1.48u_{C2} = 1.48(u_{D1} + \Delta u) \tag{15}$$

The increment $\Delta u$, as pointed out by Qin et al. (2010b), is so small that it is negligible. Thus, the displacement at the peak strength point of the first locked segment is approximately equal to that at the volume dilation point of the second locked segment. Therefore, Eq. (15) can be simplified as

$$u_{D2} = 1.48u_{D1} \tag{16}$$

Substituting Eq. (14) into Eq. (16) leads to

$$u_{D2} = 1.48^2 u_{C1} \tag{17}$$

By parity of reasoning, for a slope with $k$ locked segments, the critical instability displacement is

$$u_{Dk} = 1.48^k u_{C1} \tag{18}$$

When $k = 1$, Eq. (18) becomes Eq. (13). In this model, the critical sliding displacement of a slope with locked segments is dependent on the displacement value corresponding to the volume dilation point of the first locked segment and the number of locked segments. Hence, specific mechanical parameters of geo-materials that are difficult to measure are not required for this model.

## 5. Case studies

### 5.1 The Yanchihe rockslide

On June 3[rd], 1980, a catastrophic rockslide occurred at Yanchihe Phosphorus Mine, Hubei Province in China. Rock masses toppled from the top of the steep cliff at an elevation of 850 m, fell into the valley of 300 m depth and then slid to the north bank of the Yanchi River (Sun and Yao, 1983). The sliding body whose volume exceeded 1 million m³ formed a 500-m long and 478-m wide accumulation mass with a maximum thickness of 40 m (Figure 9). Buildings and facilities of the Mining Affair Bureau at the bottom of the valley were destroyed in 16 seconds, resulting in 284 fatalities and economic losses of 25 million RMB (Ministry of Geology and Mining Resources of the P. R. C. et al., 1991). It was one of the worst rockslide disasters recorded in China in the 20[th] Century.

The strata of the slope include phosphorite of the Doushantuo Formation (Zbd) and dolomite of the Dengying Formation (Zbdn). The bedding is monoclonal, striking NS or SSE with a dip angle of ~15° (Sun and Yao, 1983). As illustrated in the geologic profile A–A' (defined in Figure 9a and shown in Figure 10), the soft rock strata are overlain by hard rock strata, which is favorable to rockslides.

The Yanchihe rockslide was directly induced by mining operations. To prevent severe deformation in the mining area, the blasting of the roof was performed in 1979, but was not effective. However, after the blasting, six macro-cracks appeared in the sliding body. The evolution of the rockslide was primarily influenced by tension cracks 1 and 4. They propagated vertically, and approached the potential slip surface within the thin-layer dolomite. Huang (2015) argued that the rock bridge between the tips of the tension cracks and the thin-layer dolomite acted as a locked segment. According to detailed studies by Sun and Yao (1983), the sliding body was cut by crack 4 into two blocks: the primary (P) and the secondary (S) blocks (Figure 10). The eastward movement of block P was restricted by block S. Thus, the occurrence of the rockslide was mainly attributed to the macroscopic failure of the locked segment between the bottom tip of crack 4 and the potential slip surface within the dolomite layer. Although the slope had a potential bottom slip surface with a gentle inclined angle, the specific moment that induced the toppling, produced by the severe deformation in the mining area (Sun and Yao, 1983), was applied to block S. The block toppled rapidly after the locked segment failed, exhibiting brittle failure behavior. We categorize this rockslide as a case of rock bridge similar to Figure 1a, and apply our model to perform back analysis on the rockslide process.

The northward-moving process of block S can be represented by the variation of the horizontal width of crack 4 (Figure 11). A sharp increase of the width occurred on about May 23$^{rd}$, and the displacement corresponding to the volume dilation point of the locked segment ($u_C$) is ~15 cm. The predicted critical displacement at the peak strength point ($u_D$) is ~22.2 cm, which is similar to the last-observed displacement value (22.5 cm) recorded on June 2$^{nd}$, 1980.

This test result demonstrates that the evolution of the accelerating displacement stage of the slope follows Eq. (13).

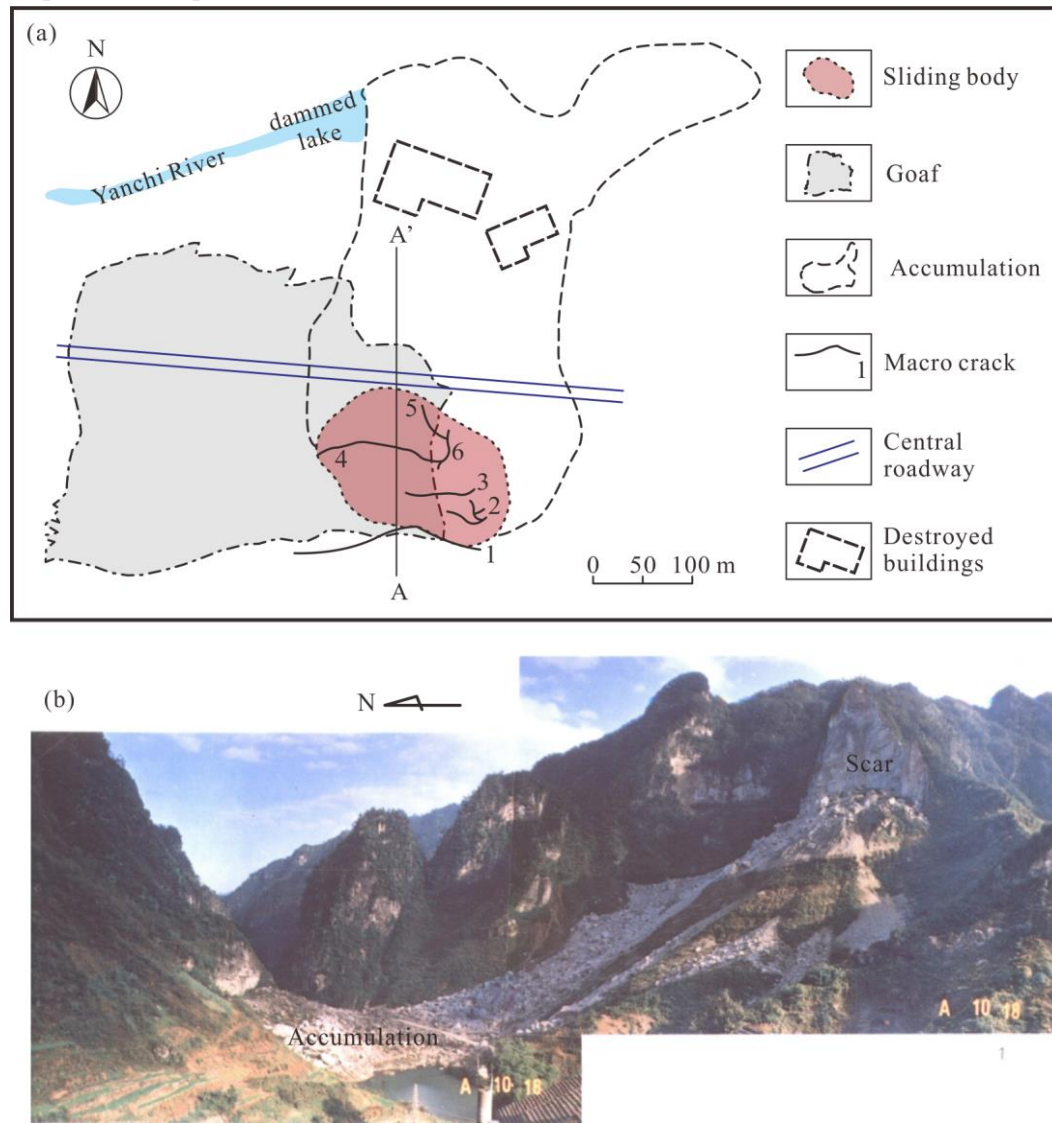

Fig. 9 The Yanchihe Mine site, Hubei Province in China; (a) ichnography of geological and engineering conditions

around the mining area (modified after Sun and Yao (1983)) and (b) the topography after the rockslide (modified after Ministry of Geology and Mining Resources of the P. R. C. et al. (1991)).

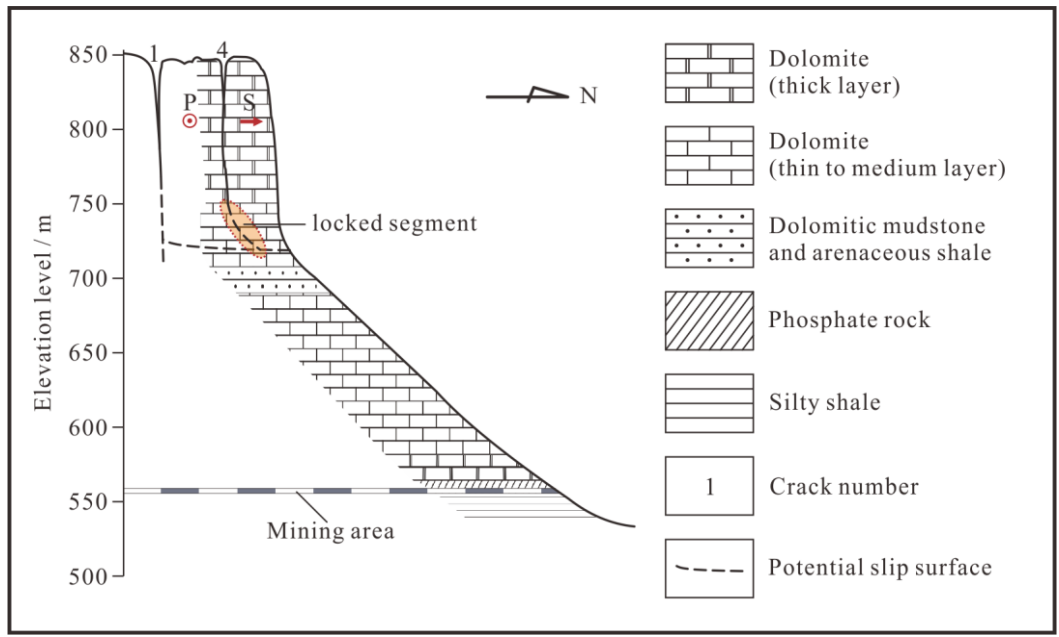

Fig. 10 Geological profile of the slope at Yanchihe Phosphorus Mine (modified after Sun and Yao (1983)). The northward and eastward movement of the two blocks, P and S, are indicated by a red arrow and a circle with a dot, respectively.

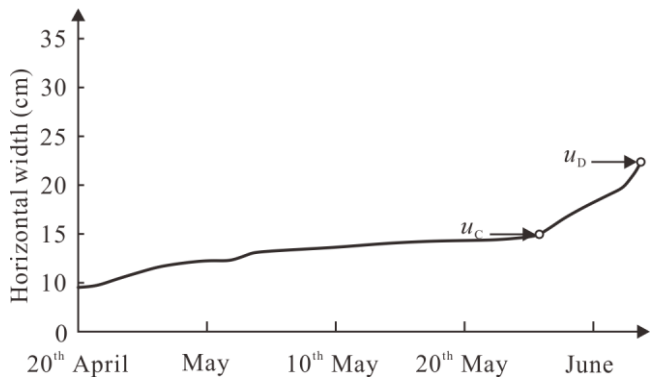

Fig. 11 Observed horizontal width of crack 4, Yanchihe Mine in 1980 (modified after Sun and Yao (1983)).

### 5.2 The wedge rockslide, Libby Dam

Libby Dam (http://www.nwd-wc.usace.army.mil/report/lib.htm) is a 128-m high concrete gravity structure in northwestern Montana, U.S.A. The lower reach of the reservoir is flanked by rock slopes. The east bank slopes are characterized by discontinuity systems, which form rock wedges plunging as steeply as 33° toward the reservoir. On January 31st, 1971, during the construction of the dam, a major portion of a rock wedge with volume of ~33,000 m³ slid from an excavated slope above the left abutment (Voight, 1979). Fortunately, this rockslide did not cause any casualties.

The bedrock of the dam and its vicinity is a thin to massive argillite of the Wallace Formation (Precambrian), with a strike N25–30°W and dip of 39–42°SW. A well-defined bedding-plane fault system exists in the east back of the Libby Dam. The fault system is intersected by several prominent sets of joints, which form many wedge-shaped "rock ribs" (Figure 12a). A wedge which slid in 1971 was bound on its upstream side by the DS+122 fault, on its downstream side by the joint A (Figure 12b and 12c), and at its rear by vertical tension cracks. Joint A, lying ENE–WSW, intersects the argillite bedding planes, which contributed to a relatively rough surface (Voight, 1979). Thus, multiple locked segments were inferred to exist along joint A (Xue et al., 2017).

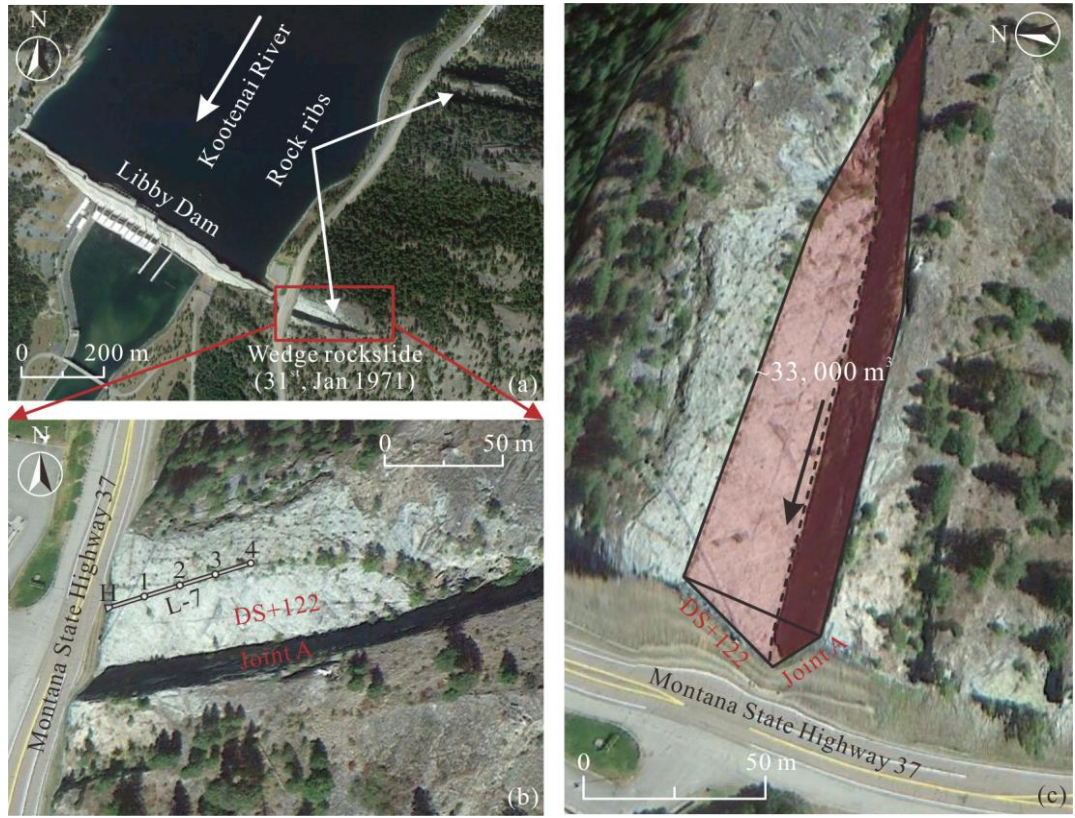

Fig. 12 Overview pictures (captured from Google Earth) of (a) the vicinity of Libby Dam and (b) the scar of the wedge rockslide (31st Jan, 1971) in the left abutment of the dam. (c) Sketch reconstructing the wedge.

Note: the location of extensometer L-7 in (b) is an approximation, compared with the accurate location of extensometer L-7 as illustrated by Voight (1979).

The movement of the sliding wedge was recorded by extensometer L-7 that was installed on the bedding plane fault DS+122 (Figure 12b) from June 1967 until pre-failure. The curves recorded by four sensors, H, 1, 2 and 3, showed the same stepped shape (Voight, 1979), indicating that the wedge moved in a cohesive manner (Xue et al., 2017). The curves are also similar to the theoretical displacement curve of a slope with multiple locked segments (Figure 8), which further supports the existence of locked segments.

Our back analysis relies on the most detailed movement record – at the floating head H of extensometer L-7 (Figure 13). The first prominent rise of the displacement (21 mm) from point (a) occurred between June and July 1967 during the blasting excavation of Montana State Highway 37 in the east abutment area (Voight, 1979). The blasting contributed to rapidly progressive failures of the first locked segment in a short time interval, leading to a rapid increase in the slope displacement. However, at that stage the volume dilation point of the locked segment had not yet been reached. After undergoing a slow long-term damage, the volume dilation point of the first locked segment was reached about in the period from May 23rd to 25th 1969, and the second dramatic increment of slope displacement (8.6 mm, from point (c) to (d)) was observed. The displacement at point (c) (22.8 mm) is defined as $u_{C1}$; using Eq. (18) we calculated $u_{D1}$, the predicted displacement corresponding to the peak strength of this locked segment, to obtain $u_{D1} = 33.7$ mm, which is quite close to the actual observation value at point (e) (33.2 mm). This result validates the inference that the slope displacement corresponding to the peak strength point of the current segment is approximately equivalent to that corresponding to the volume dilation point of the next segment. Following the same rule, one can obtain the predicted displacement value corresponding to the peak strength point of the second locked segment, $u_{D2}$, equal to 49.1 mm, which slightly exceeds the last observation (43.6 mm) recorded two weeks before the rockslide (January 18[th], 1971). Hamel (1974) argued that larger movements may have occurred on joint A in mid-January 1971. These movements were probably sufficient to open and widen certain drainage outlets along geologic discontinuities in the slope, which resulted in some drainage and relief of water pressures in the slope, thus temporarily postponing the rockslide. Therefore, it is believed that the real displacement value at instability was larger than the final observation, and was closer to our prediction.

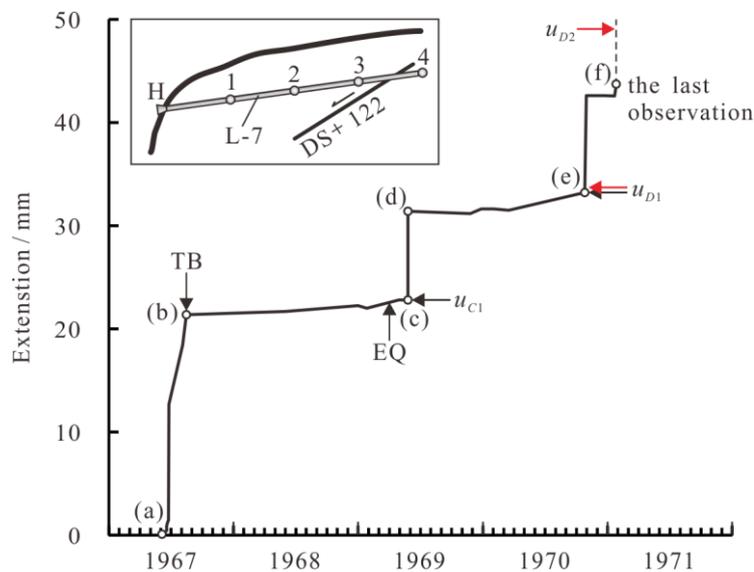

Fig. 13 Displacement records for extensometer L-7 at floating head (H) (modified after Voight (1979)). TB stands for termination of blasting. The felt earthquake of April 1[st] 1969 is noted by EQ. Predicted and observed displacement values are indicated by red and black arrows, respectively.

The above analysis using our model supports the inference that the evolution of this wedge rockslide was controlled by two locked segments. The whole progressive failure process is well described, and summarized as follows: the blasting induced the movement of the slope that had long been stable; this movement was characterized by an abrupt displacement increment induced by rapidly progressive cracking of the first locked segment with relatively low strength. After the blasting effect diminished (point (b)), the slope displacement progressed at a very slow rate due to the progressive degradation of the locked segment strength caused by environmental factors such as rainfall, snow water and earthquake. When the growing displacement value reached point (c), the volume dilation point of the first locked segment was attained, leading to the second accelerated displacement phenomenon of the slope mass. As the bearing capacity of the first lock segment decreased quickly, concurrently the second locked segment with relatively high strength bore the majority of the driving force. Therefore, the driving force and anti-sliding force came to an equilibrium state at point (d), and the movement was temporarily restricted. Once again, because of the progressive degradation of the segment strength induced by environmental factors, the slope was in a quasi-equilibrium state, exhibiting a slow displacement growth. The third rise in displacement occurred when the volume dilation point of the second locked segment was reached at point (e). Finally, when the damage of the second locked segment accumulated to reach its critical threshold, the slope instability was inevitable because all the locked segments along the potential slip surface had failed.

## 6. Discussion

There are several significant issues when applying our model in practice. To predict the critical instability displacement of a slope using Eq. (18), the displacement value corresponding to the volume dilation point of the first locked segment and the number of locked segments are required. For a slope with a locked segment, the tertiary creep can be observed from the onset of the volume dilation of the locked segment. This is a discernible precursor to the slope instability. The mechanical behavior during the tertiary creep follows a specific rule, which is the mechanical basis of predicting landslides controlled by locked segments.

Therefore, to correctly identify the accelerating displacement onset corresponding to the volume dilation point of the locked segment, a whole-process observation is strongly recommended to acquire the precise, complete movement history of the slope. However, because in-situ monitoring of slopes is usually undertaken after the distinct movement of slope is observed, this could lead to a misjudgment of the onset of tertiary creep or a serious error when estimating the critical displacement. If observations are conducted for a slope with multiple locked segments shortly after the initiation of the slope movement, the tertiary creep following long-term steady movement can usually be determined, but a displacement error preceding the observation occurs unless the displacement is represented by the accumulative width of the cracks, as in the Yanchihe rockslide. The error $\Delta$ can be calculated by Eq. (14) as

$$\Delta = \frac{u_{D1}^* - 1.48 u_{C1}^*}{0.48} \qquad (19)$$

where $u_{D1}^*$ and $u_{C1}^*$ are the in-situ measured displacement values corresponding to the peak strength point and the volume dilation point of the first locked segment, respectively. If the observation is conducted after the volume dilation point of the first locked segment is reached, the recorded data should not be used for our model to predict the landslide. Hence, observations as complete as possible is the premise of successfully predicting the slope instability by our model.

For the rock slopes in the above two cases, it is appropriate to assume that the surface displacement is proportional to the movement along the slip surface because the rock mass is almost rigid. When the slope consists of less brittle geo-materials, this hypothesis will probably be no longer valid. Thus, to improve the accuracy of the prediction when applying our model, we suggest that displacement along the potential slip surface be measured.

Moreover, if the number of locked segments is known in advance, one can judge whether the accelerated displacement phase represents the final landslide movement and correctly estimate the current stability of the slope. To this end, we suggest that future research on predicting the instability of slopes with locked segments should focus on identifying the number and location of locked segments by utilizing various geological and geophysical investigation methods.

## 7. Conclusions

A locked segment is defined as any part of a slope that governs the stability of the slope; locked segments are common in geological settings and can be classified into three categories. A physical model is proposed to predict the instability of slopes with locked segments. Applying this model to the Yanchihe and the Libby Dam rockslides, the results of analysis were in agreement with field records and enhanced our understanding of their slope instability mechanisms.

## Acknowledgments

This work was supported by the National Natural Science Foundation of China (Nos. 41572311; 41302233) and the NSFC-Henan Joint Fund (No. U1704243).